\documentclass[%
reprint,  
 amsmath,amssymb,
 aps,
]{revtex4-2}
\usepackage{enumitem}
\usepackage{graphicx}
\usepackage{dcolumn}
\usepackage{bm}
\usepackage{hyperref}
\usepackage{slashed}
\usepackage{subfloat}
\usepackage{multirow}
\usepackage[table]{xcolor}
\usepackage{pgfplots,subfigure}
\usepackage{orcidlink}
\usepackage{float}
\usepackage{geometry}
\usepackage[table]{xcolor} 
\usepackage{booktabs}       
\usepackage{array}          
\usepackage{hyperref}
\usepgfplotslibrary{fillbetween}
\usepackage{colortbl,xcolor} 
\usepackage{tikz}
\usetikzlibrary{arrows.meta}

\newcolumntype{C}{>{\centering\arraybackslash}m{4cm}} 
\definecolor{acsblue}{RGB}{17,76,139}

\begin{document}

\fontsize{7.6}{8.6}\selectfont

\title{Spectroscopic structure of Non-Hermitian $\mathcal{PT}$-Symmetric Klein--Gordon Fields in a Magnetized Cosmic String Spacetime}




\author{Omar Mustafa
}
\email{omar.mustafa@emu.edu.tr (Corresponding Author)}
\affiliation{Department of Physics, Eastern Mediterranean University, 99628, G. Magusa, north Cyprus, Mersin 10 - Türkiye}

\author{Abdullah Guvendi 
}
\email{abdullah.guvendi@erzurum.edu.tr  }
\affiliation{Department of Basic Sciences, Erzurum Technical University, 25050, Erzurum, Türkiye}


\begin{abstract}
{\fontsize{8}{8}\selectfont \setlength{\parindent}{0pt}
We investigate non-Hermitian $\mathcal{PT}$-symmetric Klein--Gordon (KG) fields in a magnetized cosmic string spacetime. A complex non-minimal scalar interaction, $\mathcal{F}_{\mu}=(0,\mathcal{F}_{r},0,0)$ with $\mathcal{F}_{r}=\omega r+ib\in\mathbb{C}$, indulged with a magnetic charge $q=ie$, is shown to introduce an effective $\mathcal{PT}$-symmetric Klein--Gordon oscillator supplemented by complex Coulombic and linear interactions. The resulting radial equation is shown to be conditionally exactly solvable using a biconfluent Heun series-to-polynomial approach. To observe non-Hermitian \(\mathcal{PT}\) symmetrization, we compare with the Hermitian counterpart by mapping $e=-i\tilde e \Rightarrow q\in \mathbb R$ and $b=-i\tilde b \Rightarrow \mathcal F_r\in \mathbb R$. In the limiting case $\omega=0$, the system is shown to reduce to an exactly solvable non-Hermitian \(\mathcal{PT}\) symmetric  Coulomb-type KG equation using confluent hypergeometric series/polynomials. Hereby, for both \(\mathcal{PT}\) symmetric models considered, we show that non-Hermitian \(\mathcal{PT}\) symmetrization introduces an upper limit for the allowed energies, unlike the corresponding Hermitian cases. These results provide an analytically tractable framework for exploring non-Hermitian relativistic quantum fields in curved spacetimes and demonstrate the role of $\mathcal{PT}$-symmetry in regulating the allowed spectrum.}
\end{abstract}


\maketitle


\section{Introduction} \label{sec:1}

\setlength{\parindent}{0pt}

Despite the absence of the conventional Hermiticity requirement, non-Hermitian Hamiltonians possessing $\mathcal{PT}$ symmetry (parity-time symmetry) can exhibit entirely real spectra within the unbroken $\mathcal{PT}$-symmetric phase. This remarkable property, first revealed in the seminal work of Bender and Boettcher in 1998 \cite{PTS1}, originates from the invariance of the Hamiltonian under the combined action of parity and time-reversal transformations, where $\mathcal{P}$ and $\mathcal{T}$ denote the parity and time-reversal operators, respectively. Consequently, the conventional Hamiltonian Hermiticity condition $H=H^\dagger$ is replaced by the mere \(\mathcal{PT}\) symmetry requirement  $H=H^\ddagger=\mathcal{PT}H\mathcal{PT}$ \cite{PTS2}. Since then, $\mathcal{PT}$ symmetry has stimulated extensive theoretical and experimental investigations in various areas of physics (see, for example, \cite{PTS3,PTS4,PTS5,PTS6,PTS7,PTS8,PTS81,PTS82} and references therein). The implications of the accelerating theoretical developments of $\mathcal{PT}$-symmetric systems have been the subject of considerable discussions and debates for many years \cite{PTS2,PTS4,PTS6}, until the recent discoveries and experimental demonstrations of $\mathcal{PT}$ symmetry in optical systems \cite{PTS9,PTS10,PTS11,PTS12,PTS13,PTS14}. These developments have opened new directions in non-Hermitian physics, including optical lattices, lasers, waveguides, superconducting systems, and electromagnetic resonators; comprehensive discussions can be found in \cite{PTS3,PTS8}.

In the present study, we investigate non-Hermitian $\mathcal{PT}$-symmetric Klein--Gordon (KG) fields in a static cosmic string spacetime in the presence of a magnetic interaction. To construct an effective $\mathcal{PT}$-symmetric coupling, we consider the analytic continuation of the electromagnetic charge, $e\rightarrow ie$, which transforms the interaction term $eA_{\mu}$ into a complex-valued non-Hermitian contribution while preserving the required $\mathcal{PT}$ symmetry properties \cite{PTS3,PTS13,PTS14,PTS15}. Such complexifications of coupling parameters, which may interchange the effective attractive and repulsive character of interactions, are closely related to earlier discussions on non-Hermitian formulations of quantum theories, including Dyson's analysis of the instability of perturbative expansions in quantum electrodynamics (QED) \cite{PTS16}. Furthermore, non-Hermitian extensions of quantum gravity theories may provide an alternative framework for exploring unconventional structures and their feasible implications \cite{PTS3,PTS15,PTS161,PTS162,PTS163,PTS164}. To the best of our knowledge, the specific construction of a $\mathcal{PT}$-symmetric Klein--Gordon field with non-minimal coupling in a cosmic string spacetime and magnetic background has not been previously explored.

On the other hand, cosmic strings are stable one-dimensional topological defects that may have been generated during symmetry-breaking phase transitions in the early universe. They represent non-trivial configurations of spacetime associated with the breakdown of cylindrical symmetry and constitute one of the most intriguing relics predicted in the context of primordial cosmology \cite{CR1,CR2,CR020,CR021,CR3,CR4,CR5,CR6,CR7}. The corresponding static cylindrically symmetric spacetime metric is given by
\begin{equation}
ds^{2}=-dt^{2}+ dr^{2}+\tilde\alpha ^{2}\,r^{2}d\varphi ^{2}+dz^{2}.
\label{1.1}
\end{equation}
Here, $0<\tilde\alpha^2=1-\eta/2\pi<1$ denotes the constant conical parameter related to the deficit angle of the cosmic string spacetime, $\eta$ relates with the linear mass density of the cosmic string. The corresponding metric tensor and its inverse are therefore expressed as
\begin{equation}
\begin{split}
g_{\mu \nu }&=diag\left( -1,\,1,\,\tilde\alpha ^{2}\,r^{2},\,1\right)
\Rightarrow \det \left( g_{\mu \nu }\right) =-\tilde\alpha ^{2}\,r^{2},\\
g^{\mu \nu }&=diag\left( -1,1,\frac{1}{\tilde\alpha ^{2}\,r^{2}},1\right).
\end{split}
\label{1.2}
\end{equation}
Within this non-trivial curved spacetime background, we consider the non-minimally coupled KG field equation
\begin{equation}
\frac{1}{\sqrt{-g}}\tilde{D}^+_{\mu }\left( \sqrt{-g}g^{\mu \nu }\tilde{D}^-_{\nu }\right) \Psi =m_\circ^{2}\Psi .
\label{1.3}
\end{equation}
where $\tilde{D}^{\pm}_{\mu}=D_{\mu}\pm\mathcal{F}_{\mu}$, $D_{\mu}$ is the gauge-covariant derivative defined by $D_{\mu}=\partial_{\mu}-iqA_{\mu}$, $m_\circ$ denotes the rest mass energy (i.e., \(m_\circ\equiv m_\circ c^2\)) of the KG field, and $\mathcal{F}_{\mu}$ represents a non-minimal coupling vector commonly employed in the construction of KG-oscillator models \cite{CR7.1,CR7.2,CR7.3,CR7.4,CR7.5,CR7.6,CR7.7,CR7.8}. Furthermore, we consider an azimuthal electromagnetic potential together with a complex non-minimal radial coupling to generate a non-Hermitian $\mathcal{PT}$-symmetric extension of the Klein--Gordon field. The resulting interaction combines the effects of the magnetic background and the complex scalar coupling, producing an effective radial potential whose spectral properties are analyzed below.

In the current proposal, we investigate, in Section II, some non-trivial non-Hermitian \(\mathcal{PT}\) symmetric KG-fields in cosmic string spacetime in a magnetic field. In the process, we consider (in section II) a non-Hermitian structure of a non-minimally coupled scalar KG field \(\mathcal F_\mu=(0,\mathcal F_r,0,0);\,\mathcal F_r=\omega r+ib \in \mathbb C, \) with a magnetic charge \(q=ie\in \mathbb{C}\), and show that it yields an effective non-Hermitian \(\mathcal{PT}\) symmetric KG oscillator perturbed by Coulombic and linear interactions. The mathematical structure of such a quantum mechanical system only allows conditional exact solvability via the biconfluent Heun series/polynomials \(H_B(\alpha,\beta,\gamma,\delta,z)\). To observe the effect of non-Hermitian \(\mathcal{PT}\) symmetrization, we  compared the results with the Hermitian counterpart by mapping \(e=-i\tilde e\to q\in\mathbb{R}\) and \(b=-i\tilde b\to \mathcal F_r\in\mathbb R\) .  We also consider the limiting case \(\omega=0\)  (i.e. a non-Hermitian constant KG-field \(\mathcal F_r=ib\) in cosmic string spacetime in a magnetic field) which results in a non-Hermitian \(\mathcal{PT}\) symmetric KG-Coulombic field. The mathematical structure of which allows for exact solvability using hypergeometric series/polynomials solutions. 
Hereby, both non-Hermitian \(\mathcal{PT}\) symmetric models provide upper limits for the allowed energies. Our concluding remarks are given in section III.

\section{Non-Hermitian \(\mathcal{PT}\)-symmetric KG-fields in cosmic string spacetime in a magnetic field} 
\label{sec:2}

We now investigate the relativistic quantum dynamics of a non-Hermitian $\mathcal{PT}$-symmetric KG field propagating in a static cosmic string spacetime in the presence of an external magnetic field. The cosmic string background introduces the conical parameter $\tilde\alpha$, which modifies the angular sector of the Klein--Gordon equation and consequently affects the radial spectrum. In addition to these purely geometrical effects, the presence of a magnetic field and a non-minimal coupling introduces further interaction mechanisms that can generate non-trivial effective potentials in the radial dynamics. In the cosmic string spacetime background (\ref{1.1}), a KG-particle with charge $q$ interacting with the electromagnetic four-potential $A_\mu=(0,0,A_\varphi,0)$, where $A_\varphi=B_\circ r/2$, is governed by the non-minimally coupled KG equation (\ref{1.2}), which takes the explicit form
\begin{equation}
 \frac{1}{\sqrt{-g}}\left(
D_{\mu }+\mathcal{F}_{\mu }\right) \sqrt{-g}g^{\mu \nu }\left( D_{\nu }-
\mathcal{F}_{\nu }\right) \Psi =m_\circ^{2}\Psi .
\label{2.1}
\end{equation}
Here, the generalized covariant derivative structure simultaneously incorporates the gauge interaction and the non-minimal coupling through the fields $D_\mu$ and $\mathcal F_\mu$, respectively. The latter coupling plays a fundamental role in the construction of KG oscillator models by introducing an effective position-dependent interaction while preserving the relativistic character of the field equation. In the present framework, this mechanism is further generalized by allowing the coupling function to acquire a complex structure, thereby providing a natural setting for investigating non-Hermitian but $\mathcal{PT}$-symmetric relativistic systems. The stationarity of the cosmic string spacetime and its invariance under translations along the string axis and rotations around it allow the separation of variables through the eigenfunction ansatz
\begin{equation}
\Psi \left( t,r,\varphi ,z\right) =\exp \left( i\left[ m \varphi
+kz-Et\right] \right) \phi \left( r\right) ,
\label{2.2}
\end{equation}
where $E$ denotes the energy eigenvalue, $k$ is the momentum along the $z$-direction, and $m=0,\pm1,\pm2,\cdots$ is the magnetic quantum number associated with the azimuthal motion. Substituting Eq. (\ref{2.2}) into Eq. (\ref{2.1}), and taking into account the non-trivial determinant and inverse metric components of the cosmic string geometry, one obtains the effective radial KG equation
\begin{equation}
\left[
\partial_r^2+\frac{1}{r}\partial_r
+\left(E^2-m_\circ^2-k^2\right)
-\tilde{\mathcal F}(r)
-\frac{\left(m-qA_{\varphi}\right)^2}{\tilde{\alpha}^{2}r^{2}}
\right]\phi(r)=0 .
\label{2.3}
\end{equation}
Equation (\ref{2.3}) contains the relativistic energy contribution, the radial kinetic operator, and the effective angular interaction modified by the cosmic string deficit and magnetic field. In particular, the cosmic string parameter $\tilde\alpha$ modifies the effective angular momentum barrier, demonstrating the sensitivity of the quantum spectrum to the global topology of the background spacetime. The non-minimal interaction contribution is summarized by the effective radial function
\begin{equation}
\tilde{\mathcal F}\left(r\right) =\mathcal{F}_{r}^{\prime }+\frac{\mathcal{F}_{r}}{r}+%
\mathcal{F}_{r}^{2}.
\label{2.4}
\end{equation}
Unlike the conventional Hermitian KG oscillator, however, we shall consider a complex deformation of the coupling field in order to explore the consequences of non-Hermitian $\mathcal{PT}$ symmetry in a curved spacetime setting. To this end, we introduce the complex magnetic charge prescription
\cite{PTS3}
\[
q=ie\Rightarrow qA_\varphi=ieA_\varphi=ieB_\circ r/2,
\]
and choose the non-minimal coupling field in the form
\[
\mathcal{F}_r=\omega r+ib .
\]
The imaginary magnetic charge and the complex scalar coupling generate compensating non-Hermitian contributions that preserve the $\mathcal{PT}$ symmetry of the resulting effective interaction. Consequently, the radial KG equation becomes
\begin{equation}
\begin{split}
&\left\{ \partial _{r}^{2}+\frac{1}{r}\partial _{r}
-\frac{\tilde{m}^{2}}{r^{2}}-{\omega}^{2}r^{2}+i\frac{\left(\tilde{m}\tilde{B} -b\right)}{r}
-2i\omega b r+\mathcal{E}^2 \right\} \phi \left( r\right) =0 .
\end{split}
\label{2.5}
\end{equation}
Equation (\ref{2.5}) defines the non-Hermitian $\mathcal{PT}$-symmetric radial oscillator problem considered below, where the harmonic confinement is supplemented by complex Coulombic and linear contributions. Since the physical radial coordinate is restricted to the half-line, $r\in[0,\infty)$, the implementation of $\mathcal{PT}$ symmetry requires a generalized interpretation of the parity operation. In the present radial formulation, the parity transformation is understood as the reflection of the analytically continued coordinate associated with the equivalent one-dimensional representation of the radial problem, while the time-reversal operation acts through complex conjugation, $i\rightarrow -i$. Within this analytically extended framework, the effective radial potential satisfies the formal $\mathcal{PT}$-symmetry condition
\[V_{\rm eff}(r)=V_{\rm eff}^{*}(-r).\]
The physical radial states are then obtained by restricting the solutions to the original domain $r\geq0$ and imposing the appropriate regularity condition at the origin. Therefore, Eq. (\ref{2.5}) may be regarded as a radial realization of a non-Hermitian $\mathcal{PT}$-symmetric problem, where the spectral structure is determined by the complex coupling parameters, the admissible boundary conditions, and the cosmic string geometry. For convenience, we introduce the effective energy parameter and the rescaled quantities
\begin{equation}
\begin{split}
\mathcal{E}^2 &={E}^2+b^2+\frac{\tilde{B}^2}{4}
-\left(k^2+m_\circ^2+2\omega\right),\\
\tilde{B}&=\frac{eB_\circ}{\tilde\alpha}, 
\qquad 
\tilde{m}=\frac{m}{\tilde\alpha}.
\end{split}
\label{2.6}
\end{equation}
The resulting equation (\ref{2.5}) constitutes the fundamental spectral problem investigated in this work. It provides a relativistic realization of a non-Hermitian $\mathcal{PT}$-symmetric Hamiltonian in a curved spacetime background, where the effects of the cosmic string geometry, electromagnetic interaction, and complex non-minimal coupling are incorporated into the radial dynamics. In the following sections, we use this effective Hamiltonian structure to identify analytically solvable sectors and examine how the $\mathcal{PT}$-symmetric deformation influences the admissible energy spectrum and the corresponding spectral constraints.

\subsection{KG-oscillators indulged with non-Hermitian \(\mathcal{PT}\) symmetric Coulombic and linear perturbations}

To explicitly expose the underlying one-dimensional Schr\"odinger-like structure, we introduce the radial transformation
\[
\phi(r)=\frac{R(r)}{\sqrt{r}},
\]
which removes the first-order derivative term and yields the non-Hermitian \(\mathcal{PT}\)-symmetric radial Klein--Gordon oscillator equation
\begin{equation}
\left[
\partial_r^2-\frac{\tilde m^2-\frac14}{r^2}-\omega^2 r^2
+i\frac{\tilde m\tilde B-b}{r}
-2ib\omega r+\mathcal E^2
\right]R(r)=0 .
\label{2.7}
\end{equation}
The inverse-square contribution determines the short-distance behavior, whereas the oscillator and complex linear terms govern the asymptotic confinement. The resulting effective potential is invariant under the combined parity-time transformation and therefore realizes a non-Hermitian \(\mathcal{PT}\)-symmetric extension of the Klein--Gordon oscillator. To obtain the exact analytical structure, we factorize the radial function as
\[
R(r)=r^{|\tilde m|+\frac12}\exp\left(-ibr-\frac{\omega r^2}{2}\right)H(r),
\]
where the first factor ensures regularity at the origin and the exponential factor guarantees the asymptotic normalizability of the radial state. With the dimensionless variable \(z=\sqrt{\omega}r\), the remaining function satisfies the canonical biconfluent Heun equation
\begin{equation}
zH''+\left[(\alpha+1)-\beta z-2z^2\right]H'
-\left[\left(\alpha+2-\gamma\right)z+\frac{\delta+\beta(\alpha+1)}{2}\right]H=0 ,
\label{2.8}
\end{equation}
where the Heun parameters are defined by
\begin{equation}
\alpha=2|\tilde m|,\qquad
\beta=\frac{2ib}{\sqrt{\omega}},\qquad
\gamma=\frac{\mathcal E^2-b^2}{\omega},\qquad
\delta=\frac{2i(b-\tilde m\tilde B)}{\sqrt{\omega}}.
\end{equation}
The regular solution of (\ref{2.8}) is represented by the biconfluent Heun expansion
\begin{equation}
H(z)=H_B(\alpha,\beta,\gamma,\delta,z)
=\sum_{j=0}^{\infty}C_jz^j .
\label{2.81}
\end{equation}
Substitution of (\ref{2.81}) into (\ref{2.8}) produces the initial coefficient relation
\begin{equation}
C_1=\frac{\delta+\beta(\alpha+1)}{2(\alpha+1)}C_0
=\frac{\delta+\beta(\alpha+1)}{2(\alpha+1)},
\qquad C_0=1 ,
\label{2.9}
\end{equation}
together with the three-term recurrence relation
\begin{equation}
\begin{split}
C_{j+2}(j+2)(j+\alpha+2)
&-C_{j+1}\left[\beta(j+1)+\frac{\delta+\beta(\alpha+1)}{2}\right]\\
&=C_j\left[2j+\alpha+2-\gamma\right],
\qquad j\geq0.\label{2.10}  
\end{split}
\end{equation}
The existence of normalizable bound states requires the infinite biconfluent Heun series to terminate into a finite polynomial. Therefore, following the standard polynomial truncation procedure, we impose
\[
\forall j=n+1,\qquad C_{n+1}\neq0,\qquad C_{n+2}=C_{n+3}=0 ,
\]
where the condition \(C_{n+2}=0\) determines the admissible parameter correlation as discussed by Ronveaux \cite{PTS17}. The first truncation condition gives
\begin{equation}
C_{n+1}\left[2(n+1)+(\alpha+2-\gamma)\right]=0
\Longrightarrow
\gamma=2(n+2)+\alpha ,
\quad n=0,1,2,\cdots 
\label{2.11}
\end{equation}
which leads to the relativistic energy spectrum
\begin{equation}
E=\pm\sqrt{
2\omega(n+|\tilde m|+3)+k^2+m_\circ^2
-\frac{e^2B_\circ^2}{4\tilde\alpha^2}}.
\label{2.12}
\end{equation}
The condition (\ref{2.11}) reproduces the standard biconfluent Heun polynomial constraint reported by Ronveaux \cite{PTS17}. Instead of treating \(\delta\) as the root of \(C_{n+2}=0\), we employ the equivalent requirement that the coefficient multiplying \(C_{n+2}\) vanishes, following the procedure adopted in related Heun-type quasi-exactly solvable problems \cite{PTS18,PTS19,PTS20,PTS21,PTS22}. This yields
\begin{equation}
\beta(n+2)+\frac12(\delta+\beta+\alpha\beta)=0
\quad\Longrightarrow\quad
m=\frac{2\tilde\alpha^2b}{eB_\circ}
\left(n+|\tilde m|+3\right).
\label{2.13}
\end{equation}
Equation (\ref{2.13}) represents the central signature of conditional exact solvability \cite{PTS23}. It establishes a non-trivial correlation among the cosmic string topology, magnetic field strength, magnetic quantum number, radial excitation number, particle charge, and the imaginary contribution of the non-minimal scalar interaction,
\[
\Im\mathcal F_r=b,
\qquad
\mathcal F_r=\omega r+ib .
\]
Therefore, the allowed spectrum does not exist for arbitrary values of the physical parameters but only within a restricted parameter subspace determined by the Heun polynomial constraint. The correlation (\ref{2.13}) can equivalently be written as
\begin{equation}
2(n+|\tilde m|+3)=\frac{meB_\circ}{\tilde\alpha^2b}\geq6 .
\end{equation}
This immediately implies that the \(m=0\) sector cannot satisfy the polynomial truncation requirement for \(n\geq0\). Consequently, the absence of the zero magnetic quantum number state emerges dynamically from the non-Hermitian \(\mathcal{PT}\)-symmetric Heun structure rather than from any externally imposed restriction. Finally, inserting the conditional solvability relation (\ref{2.13}) into the energy spectrum (\ref{2.12}) gives
\begin{equation}
E=\pm\sqrt{
\frac{m\omega eB_\circ}{\tilde\alpha^2b}
+k^2+m_\circ^2
-\frac{e^2B_\circ^2}{4\tilde\alpha^2}},
\qquad m=\pm1,\pm2,\cdots .
\label{2.14}
\end{equation}
Thus, the non-Hermitian \(\mathcal{PT}\)-symmetric deformation does not merely shift the Klein--Gordon oscillator spectrum; it imposes an intrinsic spectral selection rule through the Heun polynomial condition. Thus, the Heun constraint introduces a parameter-dependent selection rule for the admissible spectral sectors.

\subsection{Hermitian KG-fields in cosmic string spacetime in a magnetic field}
\label{subsec:II.2}

\setlength{\parindent}{0pt}

To identify the effects introduced by the non-Hermitian deformation, we consider the Hermitian counterpart obtained by replacing the complex interaction parameters with real quantities,
\[
\mathcal F_r=\omega r+\tilde b\in\mathbb R,
\qquad 
q=\tilde e\in\mathbb R .
\]
This configuration follows from the analytic continuation
\[
b=-i\tilde b,\qquad e=-i\tilde e ,
\]
which removes the imaginary components of the scalar and electromagnetic couplings. The resulting radial equation preserves the same biconfluent Heun structure as the non-Hermitian system; therefore, the polynomial truncation mechanism and the associated conditional exact solvability remain unchanged. However, the spectral contribution of the interaction terms is modified due to the transition from complex to real couplings. Applying this continuation to Eq. (\ref{2.12}) yields the Hermitian relativistic energy spectrum
\begin{equation}
E=\pm\sqrt{2\omega(n+|\tilde m|+3)+k^2+m_\circ^2+\frac{\tilde B^2}{4}},
    \qquad n=0,1,2,\cdots ,
\label{2.15}
\end{equation}
where the two branches correspond to the particle and antiparticle sectors of the Klein--Gordon field. The polynomial truncation condition of the biconfluent Heun function leads to the parameter constraint
\begin{equation}
m=\frac{2\tilde\alpha^2\tilde b} {\tilde e B_\circ}\left( n+\frac{|m|}{\alpha}+3 \right).
\label{2.16}
\end{equation}
The relation (\ref{2.16}) shows that the conditional exact solvability is an intrinsic property of the Heun structure of the relativistic oscillator in the cosmic string background and does not originate exclusively from the non-Hermitian deformation. The latter modifies the admissible parameter domain and the spectral organization rather than generating the polynomial solvability itself. Substituting Eq. (\ref{2.16}) into Eq. (\ref{2.15}), the spectrum of the conditionally solvable sector becomes
\begin{equation}
E=\pm\sqrt{\frac{m\omega \tilde e B_\circ}{\tilde\alpha^2\tilde b}+k^2+m_\circ^2+\frac{\tilde B^2}{4}},
\qquad m=\pm1,\pm2,\cdots .
\label{2.161}
\end{equation}
An important consequence of this representation is that the radial quantum number is no longer explicitly present in the energy expression, since its dependence is absorbed into the constraint (\ref{2.16}). Thus, the admissible radial states are determined by the parameter correlation required for Heun polynomial termination rather than by an independent spectral quantum number. The comparison with the non-Hermitian realization highlights the effect of the complex deformation. In the Hermitian case, the magnetic contribution appears as the positive term \(\tilde B^2/4\), whereas the non-Hermitian continuation produces the opposite sign. Therefore, the transition to a $\mathcal{PT}$-symmetric interaction modifies not only the energy values but also the structure of the allowed parameter space. The Hermitian system provides the reference spectrum in which the geometric and magnetic contributions remain conventional, while the non-Hermitian extension introduces an additional spectral regulation mechanism through complex coupling parameters.

\subsection{Non-Hermitian constant KG-field $\mathcal F_r=ib$ in cosmic string spacetime in a magnetic field: Effective $\mathcal{PT}$-symmetric KG-Coulombic field}
\label{sec:III-C}

\setlength{\parindent}{0pt}

The Coulombic sector is obtained by suppressing the oscillator contribution through the limit
\[\omega=0,\qquad \mathcal F_r\rightarrow ib .\]
In this case, the non-minimal interaction becomes a purely imaginary constant field, while the electromagnetic coupling remains non-Hermitian,
\[q=ie,\qquad qA_\varphi=ie\frac{B_\circ r}{2}.\]
The radial equation (\ref{2.7}) consequently reduces to
\begin{equation}
\begin{split}
 &\left\{\partial_r^2-\frac{\left(\tilde m^2-\frac14\right)}{r^2}+i\frac{\left(\tilde m\tilde B-b\right)}{r}+\tilde{\mathcal E}^{\,2}\right\}R(r)=0 \\
& \tilde{\mathcal E}^{\,2}=E^2-k^2-m_\circ^2+\frac{\tilde B^2}{4}+b^2 . \label{2.17}
\end{split}
\end{equation}
Unlike the oscillator case, Eq. (\ref{2.17}) belongs to the confluent hypergeometric class and admits an exact analytical treatment. The combined effect of the imaginary scalar coupling and magnetic interaction appears through the effective complex Coulomb strength \(i(\tilde m\tilde B-b)\). The regular solution is therefore expressed as
\begin{equation}
R(r)=r^{|\tilde m|+\frac12}e^{-i\tilde{\mathcal E}r}\,
{}_1F_1\left(
\frac{b-\tilde m\tilde B}{2\tilde{\mathcal E}}
+|\tilde m|+\frac12,
1+2|\tilde m|,
2i\tilde{\mathcal E}r
\right).
\label{2.18}
\end{equation}
The polynomial truncation of the confluent hypergeometric function requires
\begin{equation}
\frac{b-\tilde m\tilde B}{2\tilde{\mathcal E}}+|\tilde m|+\frac12=-n ,\qquad n=0,1,2,\cdots ,
\end{equation}
which gives the quantization condition
\begin{equation}
\tilde{\mathcal E}=-\frac{b-\tilde m\tilde B}{2n+2|\tilde m|+1}.
\label{2.19}
\end{equation}
Substitution of Eq. (\ref{2.19}) into the definition of \(\tilde{\mathcal E}\) yields the exact relativistic spectrum
\begin{equation}
E=\pm\sqrt{k^2+m_\circ^2-b^2-\frac{\tilde B^2}{4}+\frac{(b-\tilde m\tilde B)^2}{(2n+2|\tilde m|+1)^2}
}.
\label{2.20}
\end{equation}
The spectrum (\ref{2.20}) contains the relativistic kinetic and rest-mass contributions together with the effective interaction generated by the complex Coulomb coupling. The terms \(-b^2\), and \(-\frac{\tilde B^2}{4}\) represent the characteristic modification induced by the non-Hermitian deformation. A useful special case is obtained by removing the imaginary scalar interaction, \(b=0 \). The system then describes a Klein--Gordon particle with purely imaginary charge in the magnetic cosmic string background. The corresponding spectrum becomes
\begin{equation}
E=\pm\sqrt{k^2+m_\circ^2-\frac{\tilde B^2}{4}+\frac{\tilde m^2\tilde B^2}
{(2n+2|\tilde m|+1)^2}}.
\label{2.21}
\end{equation}
In this limit, the dependence on the magnetic quantum number appears only through \(\tilde m^2\); consequently, the transformation \(m\rightarrow -m\) leaves the spectrum invariant and the opposite angular momentum sectors remain degenerate. For comparison with the Hermitian Coulomb configuration, we perform the inverse continuation \(e\rightarrow -i\tilde e \). The energy spectrum then becomes
\begin{equation}
E=\pm\sqrt{k^2+m_\circ^2+\frac{\tilde B^2}{4}-\frac{\tilde m^2\tilde B^2}{(2n+2|\tilde m|+1)^2}}.
\label{2.22}
\end{equation}
Equations (\ref{2.21}) and (\ref{2.22}) demonstrate the different spectral signatures of the two realizations. The Hermitian continuation restores the conventional positive magnetic contribution, whereas the non-Hermitian theory introduces the opposite sign through the complex electromagnetic coupling. Therefore, the Coulombic limit provides an analytically transparent example in which the $\mathcal{PT}$-symmetric deformation modifies the relativistic spectrum through the complex interaction parameters.

\begin{figure}[ht]
\centering
\includegraphics[scale=0.58]{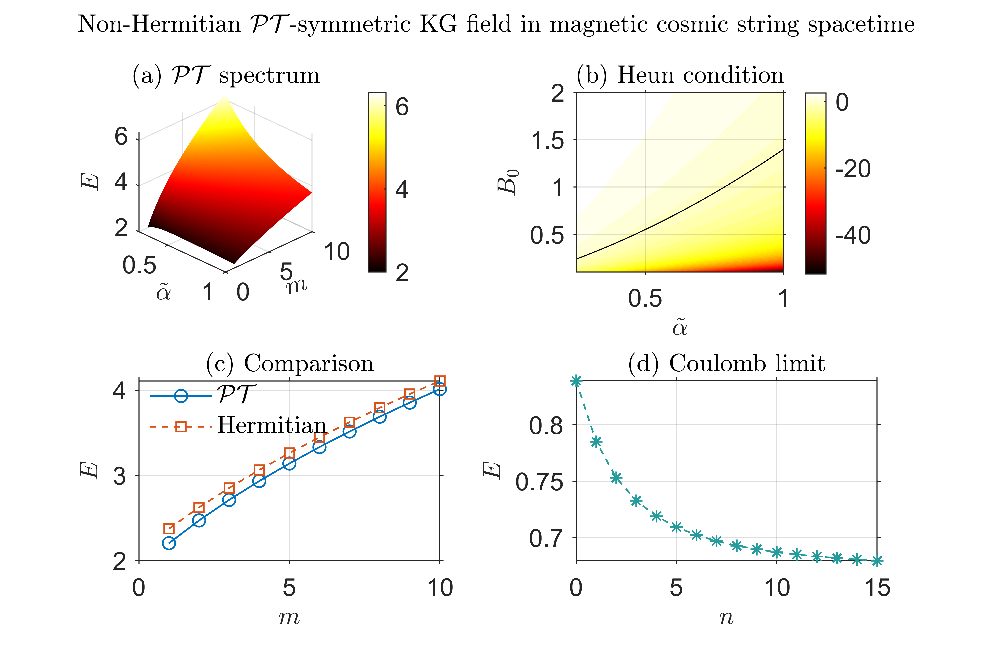}\\
\caption{\fontsize{8}{9}\selectfont Dependence of the relativistic energy spectrum of the non-Hermitian
$\mathcal{PT}$-symmetric Klein--Gordon field in the magnetized cosmic string background. Panels (a)--(d) show the variation of the energy eigenvalues with respect to the selected control parameter while keeping the remaining parameters fixed. The calculations are performed using $\tilde{\alpha}=0.9$, $m_{\circ}=1$, $k=1$, $e=1$, $\omega=1$, and $b=1$, with the magnetic quantum number and radial excitation chosen as
$m=\pm1$ and $n=0,1,2,3$, respectively. The displayed spectra correspond to the conditionally solvable sector determined by the biconfluent Heun polynomial constraint $m=2\tilde{\alpha}^{2}b(n+|\tilde m|+3)/(eB_{\circ})$. The figure demonstrates how the combined effects of the cosmic string topology, magnetic interaction, and non-Hermitian coupling modify the allowed energy levels.}
\label{fig:fig}
\end{figure}
The energy spectrum displayed in Fig.~\ref{fig:fig} exhibits the dependence of the admissible states on the cosmic string geometry, the external magnetic field, and the non-Hermitian $\mathcal{PT}$-symmetric coupling. The conical parameter $\tilde{\alpha}$ modifies the effective angular momentum contribution
through $\tilde m=m/\tilde{\alpha}$, thereby affecting the structure of the energy levels. The magnetic field contribution is modified by the imaginary charge prescription $q=ie$, which changes the sign of the corresponding magnetic term in the relativistic spectrum. In addition, the existence of bound-state solutions is restricted by the biconfluent Heun polynomial condition, which imposes a correlation among the radial quantum number, magnetic quantum number, magnetic field strength, and the imaginary component
of the non-minimal coupling. Therefore, the spectral behavior shown in Fig.~\ref{fig:fig} provides a direct consequence of the conditional exact solvability of the non-Hermitian Klein--Gordon system and demonstrates that the $\mathcal{PT}$-symmetric deformation modifies both the energy spectrum and the
allowed parameter space of normalizable states.

\section{Concluding remarks} \label{sec:4}

\setlength{\parindent}{0pt}

In this work, we have studied a non-Hermitian $\mathcal{PT}$-symmetric Klein--Gordon field in a magnetized cosmic string spacetime. By introducing the complex non-minimal coupling
\[
\mathcal{F}_{\mu}=(0,\mathcal{F}_{r},0,0),\qquad \mathcal{F}_{r}=\omega r+ib,
\]
together with the imaginary electromagnetic charge prescription $q=ie$ \cite{PTS3}, we have constructed a relativistic extension of the Klein--Gordon oscillator containing complex Coulombic and linear interactions. The resulting radial equation describes a non-Hermitian $\mathcal{PT}$-symmetric system whose spectral properties are modified by the cosmic string geometry, magnetic field, and complex coupling parameters.

\setlength{\parindent}{0pt}

The radial Klein--Gordon equation has been reduced to the biconfluent Heun form. Imposing the finiteness condition requires the infinite Heun series to terminate, resulting in a conditionally exactly solvable sector characterized by polynomial solutions \cite{PTS17,PTS18,PTS19,PTS20,PTS21,PTS22,PTS23}. The polynomial truncation condition generates a non-trivial relation among the radial quantum number, magnetic quantum number, magnetic field strength, cosmic string parameter, and the imaginary part of the non-minimal coupling. Consequently, the allowed bound states are restricted to a specific parameter domain rather than forming an unrestricted spectrum. This constraint represents a distinctive spectral feature of the non-Hermitian $\mathcal{PT}$-symmetric deformation.

\setlength{\parindent}{0pt}

To separate the effects of complex deformation from those of the conventional oscillator structure, we have analyzed the Hermitian realization obtained through the analytic continuation \(e=-i\tilde e,\quad b=-i\tilde b \). The comparison demonstrates that the biconfluent Heun solvability structure remains unchanged, while the spectral contributions associated with the magnetic and scalar couplings acquire different signatures. In particular, the imaginary electromagnetic coupling changes the magnetic contribution in the energy spectrum, leading to a different admissible spectral domain compared with the Hermitian configuration. Thus, the non-Hermitian extension modifies both the eigenvalue structure and the parameter conditions required for the existence of finite/normalizable solutions.

\setlength{\parindent}{0pt}

We have also examined the limit $\omega=0$, in which the oscillator interaction disappears and the system reduces to a non-Hermitian Coulomb-type Klein--Gordon problem. In this case, the biconfluent Heun equation is replaced by the confluent hypergeometric equation, allowing for an exact analytical determination of the energy spectrum. The resulting expression explicitly reveals the contribution of complex electromagnetic and scalar couplings and provides a simpler setting to identify the spectral consequences of the $\mathcal{PT}$-symmetric deformation.

\setlength{\parindent}{0pt}

The numerical behavior presented in Fig.~\ref{fig:fig} confirms the analytical structure of the spectrum. The variation of the energy levels with the external parameters shows that the cosmic string parameter $\tilde{\alpha}$ modifies the effective angular contribution through \(\tilde m=m/\tilde{\alpha}\), thereby changing the organization of the admissible states. The magnetic field dependence is altered by the imaginary charge prescription $q=ie$, which produces the characteristic negative magnetic contribution appearing in the non-Hermitian spectrum. Moreover, the displayed levels correspond only to the parameter sector satisfying the biconfluent Heun truncation condition,
\[
m=\frac{2\tilde{\alpha}^{2}b}{eB_\circ}\left(n+|\tilde m|+3\right),
\]
showing that the spectral curves represent the conditionally solvable states rather than an arbitrary choice of quantum numbers. Hence, the behavior illustrated in Fig.~\ref{fig:fig} provides a direct representation of the restrictions imposed by the Heun polynomial structure and the modifications introduced by the $\mathcal{PT}$-symmetric interaction.

\setlength{\parindent}{0pt}

The results demonstrate that non-Hermitian $\mathcal{PT}$-symmetric couplings can be consistently incorporated into relativistic Klein--Gordon fields in curved spacetime backgrounds. The cosmic string topology modifies the effective angular sector, while the complex interactions introduce additional constraints on the admissible spectrum. This framework provides an analytically solvable setting for investigating the influence of non-Hermitian structures on relativistic quantum fields in non-trivial geometries.

\setlength{\parindent}{0pt}

Future studies may investigate the boundaries of the unbroken $\mathcal{PT}$-symmetric regime, alternative confinement mechanisms, and extensions to more general gravitational backgrounds. Such developments may further clarify the role of non-Hermitian structures in relativistic quantum theories and curved spacetime physics \cite{PTS15,PTS161,PTS162,PTS163,PTS164}.


\end{document}